\documentclass[aps,prl,showpacs,showkeys,amsmath,amssymb,
twocolumn,
floatfix,
superscriptaddress
]{revtex4}

\usepackage{amsmath}
\usepackage{graphicx}
\usepackage{multirow}
\usepackage{longtable}
\usepackage{rotating}

\usepackage{amsmath,amstext,amsfonts,amsbsy,amssymb,amscd,bbm,epsfig}

\newcommand{\be}{\begin{equation}}
\newcommand{\ee}{\end{equation}}
\newcommand{\bea}{\begin{eqnarray}}
\newcommand{\eea}{\end{eqnarray}}

\allowdisplaybreaks[4] 

\begin{document}

\begin{titlepage}

\renewcommand{\thefootnote}{\alph{footnote}}

\vspace*{-3.cm}
\begin{flushright}
IDS-NF-037 
\end{flushright}

\vspace*{0.2cm}

\title{Physics Performance of a Low-Luminosity Low Energy Neutrino Factory}

\author{E.~Christensen}
\affiliation{Center for Neutrino Physics, Virginia Tech, Blacksburg, VA 24061, USA}
\author{P.~Coloma}
\affiliation{Center for Neutrino Physics, Virginia Tech, Blacksburg, VA 24061, USA}
\author{P.~Huber}
\affiliation{Center for Neutrino Physics, Virginia Tech, Blacksburg, VA 24061, USA}

\date{\today}

\begin{abstract}

  We investigate the minimal performance, in terms of beam luminosity
  and detector size, of a neutrino factory to achieve a competitive
  physics reach for the determination of the mass hierarchy and the
  discovery of leptonic CP violation. We find that a low luminosity of
  $2\times 10^{20}$ useful muon decays per year and 5\,GeV muon energy aimed
  at a 10\,kton magnetized liquid argon detector placed at 1300\,km
  from the source provides a good starting point.  This result relies
  on $\theta_{13}$ being large and assumes that the so-called platinum
  channel can be used effectively. We find that such a minimal
  facility would perform significantly better than phase~I of the LBNE
  project and thus could constitute a reasonable step towards a full
  neutrino factory.
\end{abstract}

\maketitle

\end{titlepage}

The recent discovery of
$\theta_{13}$~\cite{An:2012bu,Ahn:2012nd,Abe:2012tg} is a major step
towards the completion of the leptonic mixing matrix. The remaining
unknown mixing parameters, within a three neutrino framework, are the
Dirac CP-violating phase, $\delta$, and the ordering of the neutrino
mass eigenstates, $\textrm{sgn}(\Delta m_{31}^2)$. CP violation (CPV) within
the Standard Model has proved to be quite intruiging in the hadronic
sector already: even though the strong interaction seems to be
conserving CP, it is significantly violated in quark mixing.
Neutrinos now offer the third opportunity to learn more about the role
of the CP symmetry in Nature. Also, if one considers the question of
unitarity and the completeness of the three neutrino picture, the
determination of the CP phase will play a crucial role, like it did in
the quark sector.

Direct CPV in neutrino oscillations can only be
observed in appearance experiments, where the initial and final
neutrino flavors are different.  For practical reasons, this
requirement confines experiments to study
$\nu_e\leftrightarrow\nu_\mu$ and
$\bar\nu_e\leftrightarrow\bar\nu_\mu$ transitions. Conventional
neutrino beams are obtained from the decay of relativistic pions and
therefore predominantly consist of $\nu_\mu$ or $\bar\nu_\mu$,
depending on whether $\pi^+$ or $\pi^-$ are selected at the beam
source.  The current generation of experiments employing this type of
beam has a limited sensitivity to CPV, even if their results are
combined~\cite{Huber:2009cw}.

Therefore, a number of new experiments has been proposed in order to
observe CPV in the leptonic sector, see for instance
Ref.~\cite{Agarwalla:2012mz}; in the U.S. context, this proposal is
the long baseline neutrino experiment (LBNE). The first stage of the
LBNE project comprises a 700\,kW proton beam to produce pions and a
10\,kton liquid argon (LAr) detector placed at a distance $L=1300$\,km
from the source~\cite{CDR}.  The CPV discovery potential is limited
due to a lack of statistics, though.  An upgraded beam in the multi-MW
range (superbeam) would obviously yield a much better physics
potential~\cite{Akiri:2011dv}.  However, these beams are eventually
limited by intrinsic backgrounds and systematic effects: large flux
uncertainties, combined with the inability to measure the final flavor
cross sections at the near detector, introduce large systematical
errors which are very difficult to
control~\cite{Huber:2007em,Coloma:2012ji}.

For the determination of the CP phase a similar precision to that
achieved in the quark sector is only offered by a neutrino factory
(NF)~\cite{Coloma:2012wq,Coloma:2012ji}. In a NF a highly collimated
beam of muon neutrinos and electron antineutrinos is produced from
muon decays in a storage ring with long straight
sections~\cite{Geer:1997iz}. Muon decays result in a beam with equal
number of $\nu_\mu$ and $\bar\nu_e$; the CP-conjugate beam is obtained
from $\mu^+$ decays. The main observables at the NF for $\mu^-$ decay
are: $\bar\nu_e\rightarrow\bar\nu_\mu$, so-called golden channel~\cite{Cervera:2000kp}; and
the $\nu_\mu\rightarrow\nu_\mu$ disappearance channel.  At the
detector, the signal is in both cases extracted from the
charged-current interaction sample. 
Therefore, the electric charge of the muon needs to be identified in
order to disentangle the appearance and disappearance signals.  Charge
identification typically is achieved by employing a magentic field in
0.2-2\,T range. Moreover, an appropriate detector and/or muon energy
would allow to observe additional channels --
$\nu_\mu\rightarrow\nu_e$ (platinum), $\nu_\mu\rightarrow \nu_\tau$
and $\nu_e\rightarrow\nu_\tau$ (silver), see for instance
Refs.~\cite{Huber:2006wb,Donini:2002rm}.

The NF was originally proposed to operate at very high energies,
$E_\mu\sim 25-50$ GeV, and optimized under the assumption of a very
small $\theta_{13}$, $\sin^22\theta_{13}\lesssim 10^{-3}$. However, it
has recently been argued that a lower energy version would be better
optimized for the large $\theta_{13}$ scenario and technically less
demanding.  Therefore, the present NF design
parameters~\cite{NF:2011aa} are $10^{21}$ useful muon decays per
$10^7$ seconds, aimed to a 100\,kton magnetized iron detector (MIND)
placed at 2000\,km from the source, with a parent muon energy of
10\,GeV. The performance of this setup is remarkable, and clearly
superior to that of any conventional muon beam, see for instance
Ref.~\cite{Agarwalla:2012mz}. However, in order to form an intense
muon beam for acceleration and storage, muon phase space cooling is
required for this default configuration. In addition, the fact that
neutrinos in a NF are a tertiary beam implies significant proton
driver intensities; in this case, a 4\,MW proton beam together with
its associated target station. These technical challenges are to be
contrasted with the advantages of a NF -- there are no intrinsic
backgrounds and the absolute neutrino flux can be determined to much
better than 1\%.  Furthermore, the presence of both muon and electron
neutrinos in the beam does allow for a measurement of all final flavor
cross sections at the near detector.

Recently, it has been suggested that a very low energy NF, now called
nuSTORM~\cite{Kyberd:2012iz}, with greatly reduced beam power and a
muon energy around 4\,GeV, could be used for sterile neutrino searches
as well as to perform neutrino cross section measurements in the low
energy regime. Such a facility could well serve as a first stage
towards a full NF. Nevertheless, in order to do
long baseline oscillation studies, additional acceleration would be
required to achieve enough statistics at the far detector. What we
propose in this work is to use a neutrino beam which would lie in
between nuSTORM and a full NF in terms of luminosity. Specifically, we
study a configuration with a muon energy of 5\,GeV and $10^{20}$
useful muon decays per year and polarity, which already implies a
reduction of a factor of 10 over the luminosity usually considered for
NF setups. The beam energy of 5\,GeV is optimal for the considered
distance of 1\,300\,km since it balances the position of the
oscillation maximum with the peak of the event
distribution~\cite{Ballett:2012rz,Tang:2009wp}. The choice of the number of useful
muon decays is inspired by a recent study based on Fermilab's planned
Project~X accelerator complex~\cite{MAP}. Specifically, the number
we use corresponds to a proton beam of 1\,MW power at 3\,GeV and does
\emph{not} assume any muon phase space cooling. As a result, most of
the technical difficulties of a full NF can be avoided.

\begin{table}[htb]
\begin{center}
{\renewcommand{\arraystretch}{1.6}
\begin{tabular}{cc|ccccc}
& Channel 		& Effs. 		 	& Bg. Rej. 		& $\sigma (E_\nu)$ 		& $E_\nu$ (GeV) 		\\ \hline

\multirow{2}{*}{\begin{sideways} LAr \phantom{l}  \end{sideways}} 
& $\nu_\mu$  		& 80\% 				& 99.9\% 		& $0.2\sqrt{E} $ 		& $\left[ 0.5, 5 \right]$ 	\\ 
& $\nu_e$  		& 80\% 				& 99\% 		& $0.15\sqrt{E} $		& $\left[ 0.5, 5 \right]$ 	\\ \hline

\multirow{2}{*}{\begin{sideways} TASD \phantom{}  \end{sideways}} 
& $\nu_\mu$  		& 73\%-94\% 			& 99.9\% 		& $0.2\sqrt{E} $ 		& $\left[ 0.5, 5 \right]$ 	\\ 
& $\nu_e$  		& 37\%-47\% 			& 99\% 			& $0.15\sqrt{E} $		& $\left[ 0.5, 5 \right]$ 	\\ \hline

\end{tabular}}
\caption{Main details used to simulate the LAr and TASD detector
  responses. The two rows correspond to the details used for $\nu_\mu$
  and $\nu_e$ detection. The different columns indicate: signal
  efficiencies, background rejection efficiencies (Neutral Current,
  Charge misID, Flavor misID), energy resolution and neutrino energy
  range. NC backgrounds have been migrated to lower energies using
  LBNE migration matrices~\cite{Akiri:2011dv}. }
\label{tab:det}
\end{center}
\end{table}

As already mentioned, a fundamental advantage of the NF with respect
to other possible neutrino beams is the possibility to observe many
oscillation channels using the same neutrino beam. The combination of
CP- and CPT-conjugate channels not only provides crucial constraints
for the observation of leptonic CP violation and/or effects coming
from new physics. It also helps to mitigate the
effect from the matter density uncertainty~\cite{Huber:2006wb}, which
constitutes the most relevant source of systematic uncertainties at a
NF~\cite{Huber:2002mx,Coloma:2012ji}. Therefore, the simultaneous
observation of both golden and platinum channels at a NF would yield
extremely robust results, since they are CPT conjugates. However, the
platinum channel is inaccessible in a MIND, because it requires to
identify electron neutrino charged current events, and electric charge
identification is again needed in this case in order to disentangle
the $\nu_e$ appearance and $\bar\nu_e$ disappearance signals.  The
feasibility of electron charge identification was studied in
Ref.~\cite{Bross:2007ts} in the context of a low energy NF for a
totally active scintillator detector (TASD) magnetized to 0.5\,T using
a so-called magnetic cavern. It soon was speculated that a magnetized
LAr detector should be suitable as
well~\cite{Bross:2009zzb,FernandezMartinez:2010zza,Ballett:2012rz}.

\begin{figure}[t!]
  \includegraphics[height=0.8\columnwidth]{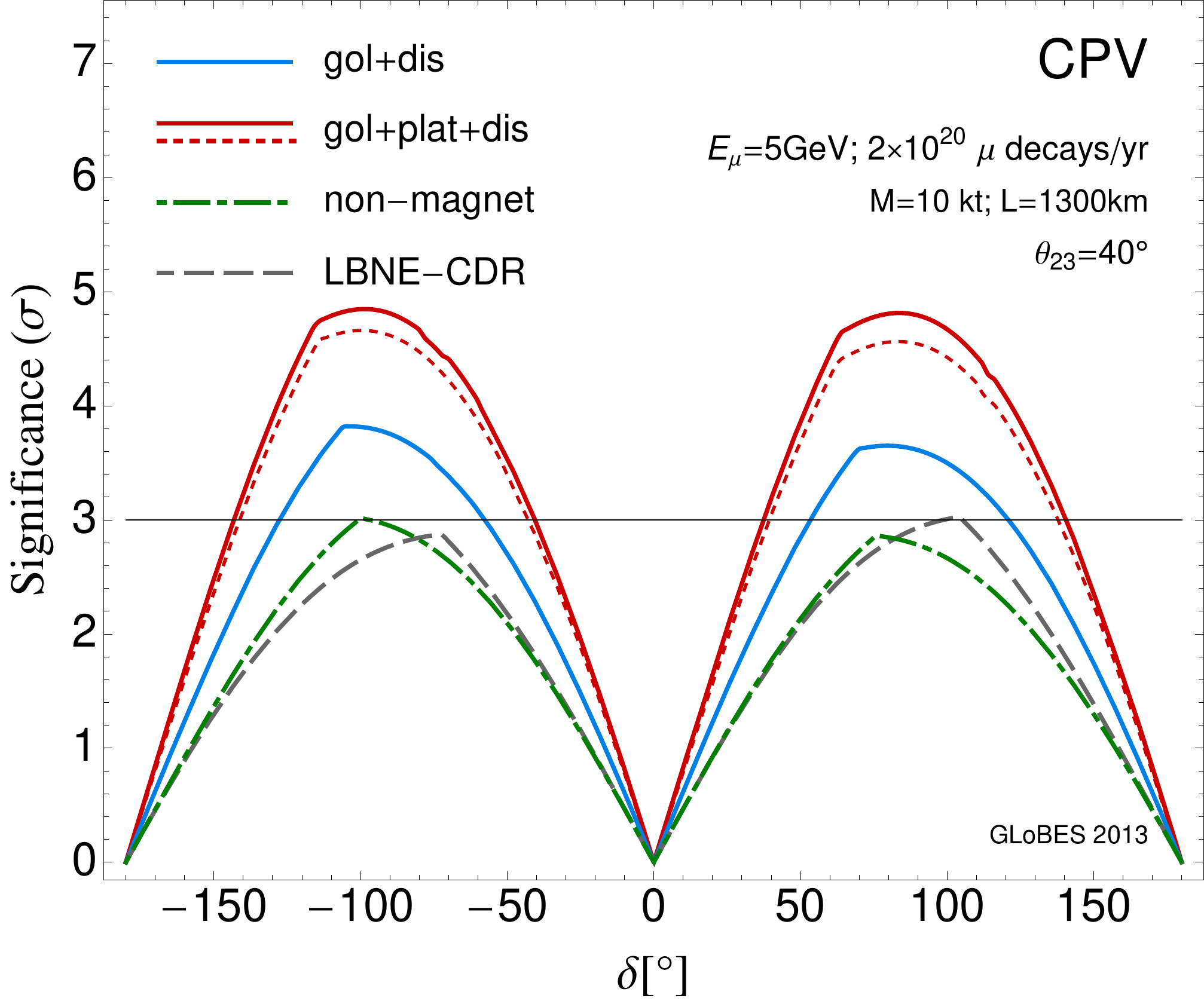} 
  \caption{ CPV discovery potential as a function of the true value of $\delta$.
    Results are shown for the combination of only the golden and $ \nu_\mu$ disappearance signals (blue lines, ``gol+dis''), as well as
    when the platinum signal is also considered (red lines,
    ``gol+dis+plat''). Solid (dotted) red lines show the results for a magnetized LAr (TASD)
    detector. Dot-dashed green lines show the results for a 10 kton non-magnetized LAr detector. For reference, the
    results for LBNE phase I are also
    shown (dashed gray lines).  }
\label{fig:disc}
\end{figure}

In this work we consider a 10\,kt magnetized LAr detector at a
distance of 1\,300\,km. This choice of detector size and baseline is
obviously inspired by LBNE: it allows to reuse the LBNE facilities to
the largest possible extent. Note, that the detector most likely will
have to be deep underground due to the large duty factor of stored
particle beams as in a NF. Tab.~\ref{tab:det} summarizes the detector
parameters used in this work. In the absence of a detailed study of
the performance of a magnetized LAr detector, we have followed
Refs.~\cite{FernandezMartinez:2010zza,Akiri:2011dv}.  The expected total event
rates for a 10 kton LAr detector are shown in
Tab.~\ref{tab:events}. Since the LAr performance is indeed uncertain,
we also evaluate sensitivities using the performance parameters of a
TASD, which are based on simulation studies~\cite{Bross:2007ts}. The
same background migration matrices and energy resolution as for the
LAr detector have been considered.  Energy dependent efficiencies for
the signal, following Ref.~\cite{FernandezMartinez:2010zza}, have been
used in this case, see Tab.~\ref{tab:det}. In addition to the
backgrounds considered in previous references, the
$\tau$-contamination~\cite{Indumathi:2009hg,Donini:2010xk,Dutta:2011mc}
has also been included in this work.  Systematic uncertainties have
been implemented as in Ref.~\cite{Coloma:2012ji}, using the default
values listed in Tab.~2 therein.

\begin{table}[htb]
\begin{center}
{\renewcommand{\arraystretch}{1.5}
\begin{tabular}{l|c|c|c}
Channel 		& \; $\nu_e\rightarrow \nu_\mu$ \; 	& \; $\nu_\mu\rightarrow \nu_e$ 	\; & \; $\nu_\mu\rightarrow \nu_\mu$  \; \\ \hline
 Signal		    &  267 & 276 & 1485	\\ 
 Background \qquad	&    7   &  73 &  17	\\ \hline
 
Channel 		&\; $\bar\nu_e\rightarrow \bar\nu_\mu$  \;	& \; $\bar\nu_\mu\rightarrow \bar\nu_e$ \; 	& \; $\bar\nu_\mu\rightarrow \bar\nu_\mu$ \; \\ \hline
 Signal 	        & 52 & 59  &  562	\\ 
 Background	&  6  &  73 &  6	\\ \hline
\end{tabular}}
\caption{Expected total number of events for the low luminosity NF
  aiming to a 10 kton LAr detector, for $\sin^22\theta_{13}=0.1$ and
  $\delta=0$.  The experiment is assumed to run with both polarities
  circulating in the ring at the same time, for 10 years. This results
  in a total of $2\times10^{21}$ muon decays in the straight sections
  of the storage ring (half per polarity). Signal and background
  rejection efficiencies are already accounted for. }
\label{tab:events}
\end{center}
\end{table}

\begin{figure*}[t!]
  \includegraphics[height=6cm]{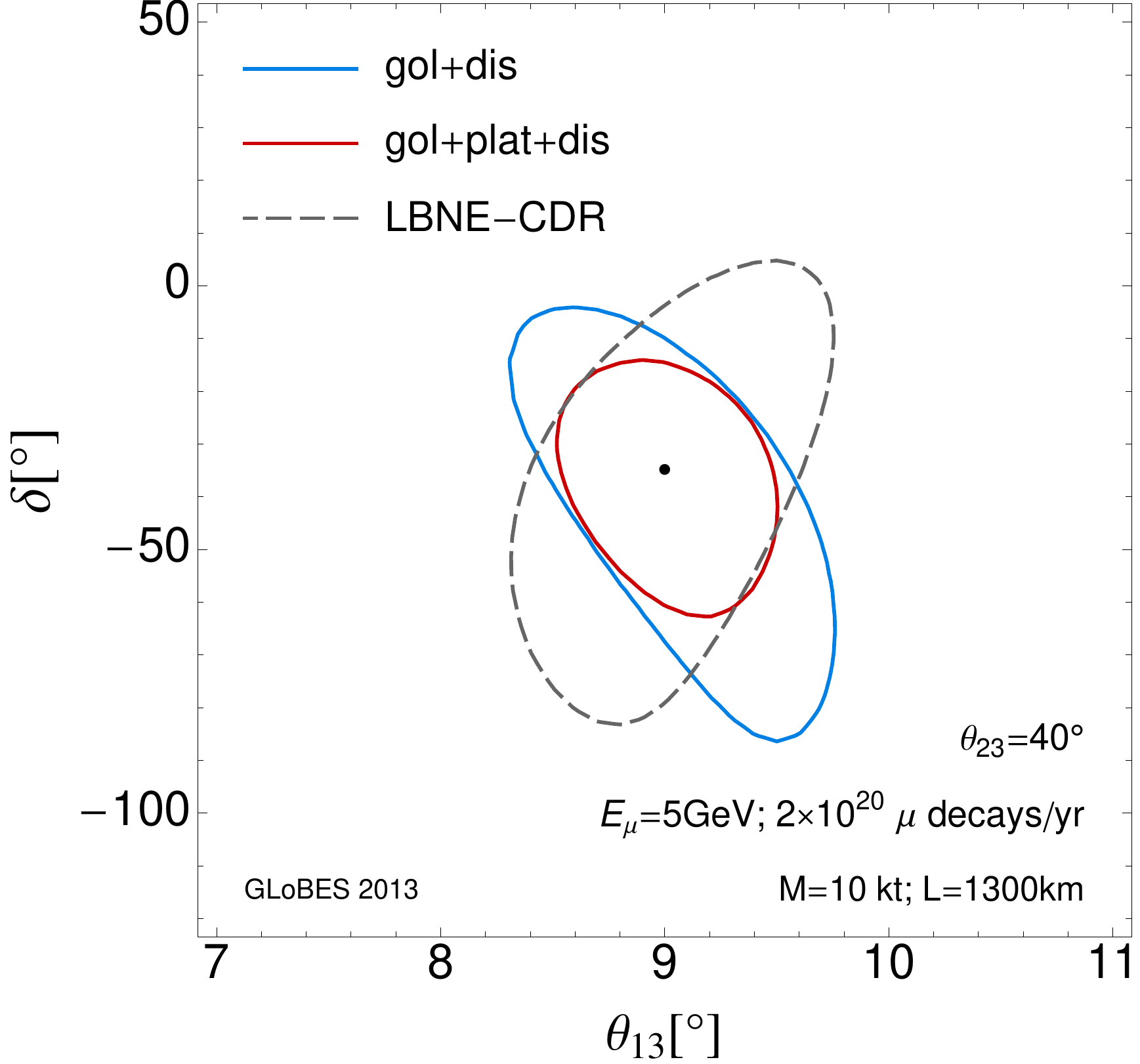}\hspace{2ex}%
  \includegraphics[height=6cm]{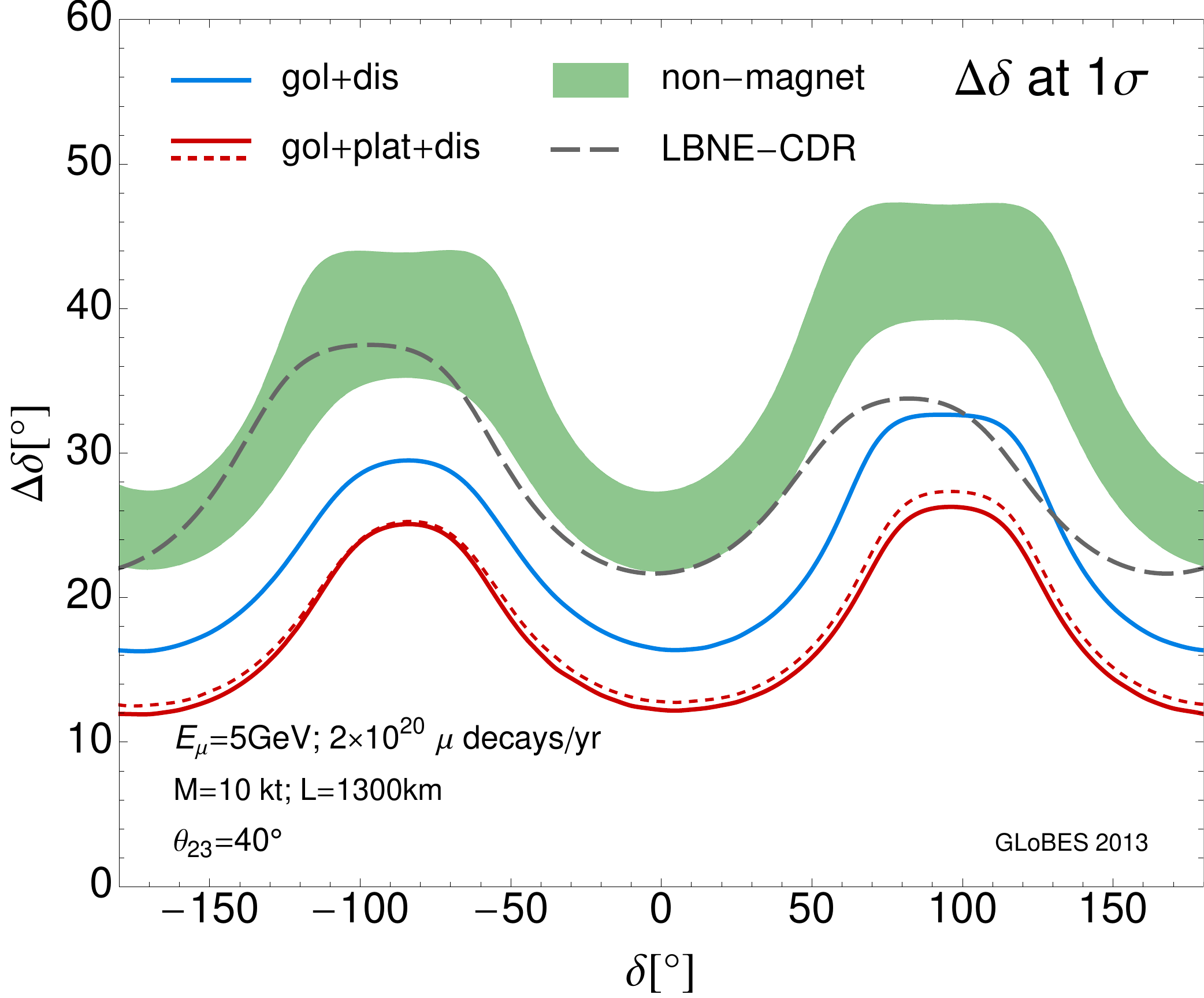}
  \caption{ Left panel: confidence region in the $\theta_{13}-\delta$
    plane for a particular point in the parameter space, at 1$\sigma$
    CL (2 d.o.f.). Right panel: precision achievable for a measurement
    in $\delta$ at the 1$\sigma$ CL (1 d.o.f.), as a function of the
    true value of $\delta$.  Results are shown when only the golden
    and disappearance channels are included in the analysis (blue
    lines, ``gol+dis''), as well as when the platinum channel is also
    considered (red lines, ``gol+dis+plat''). Solid (dotted) red lines
    show the results for a magnetized LAr (TASD) detector. The green
    bands show the physics reach for a 10 kton non-magnetized LAr
    detector, see text for details.  For reference, the results for
    LBNE phase I are also shown (dashed gray lines). }
\label{fig:prec}
\end{figure*}

Figure~\ref{fig:disc} shows the results for the CPV discovery
potential of the facility, defined as the ability of the experiment to
rule out the CP conservation hypothesis ($\delta=0,\pi$). The
statistical significance of the signal is shown as a function of the
true value of $\delta$. For reference, we also show the results for
phase I of the LBNE experiment, which has been simulated according to
the CDR from October 2012, Ref.~\cite{CDR}.  It should be noted that
for the LBNE results systematic uncertainties have been implemented as
overall normalization errors over all signal and background
contributions at once (no near detector has been simulated for this
setup). Clearly, the low energy, low luminosity NF outperforms LBNE by
a considerable margin, and the results combining only the golden and
disappearance signals are already better; as expected, if the platinum
signal is added then the performance is considerably improved. If
magnetization of a massive LAr were not possible, several methods
would in principle allow to statistically differentiate the charge of
the leptons produced at a LAr detector, see for instance
Ref.~\cite{Huber:2008yx}. Therefore, we also show in
Fig.~\ref{fig:disc} the performance of the setup using a
non-magnetized LAr detector, simulated following
Ref.~\cite{Huber:2008yx} (dot-dashed green lines). We assume that
$\nu/\bar\nu$ separation at the 90\% (70\%) for $\mu$-like ($e$-like)
events can be obtained for a non-magnetized LAr detector. Regarding
the MH discovery potential, we find that a low luminosity NF combined
with a LAr (TASD) detector can rule out the wrong hierarchy at $\sim
10\sigma$ ($8\sigma$) CL for 1 d.o.f., regardless of the true value of
$\delta$. It should be kept in mind that LBNE phase I would reach
$3\sigma$ ($5\sigma$) CL for approximately 75\% (50\%) of the values
of $\delta$~\cite{CDR}.

The left panel in Fig.~\ref{fig:prec} shows the allowed region in
$\theta_{13}$-$\delta$ plane for one particular point in the parameter
space, where the different line styles correspond to different
combinations of channels. Clearly, the addition of the platinum
channel improves the performance beyond a mere increase of statistics
-- a true synergy, whose origin is explained in
Ref.~\cite{Huber:2006wb}. The right hand panel, on the other hand,
shows the achievable precision for a measurement of $\delta$ at
1\,$\sigma$ as a function of the true value of $\delta$. Again, we
find that the low luminosity low energy NF constitutes a marked
improvement over LBNE. We also show in this case a green band, which
corresponds to the results using a 10 kton non-magnetized LAr
detector. The lower limit in the band corresponds to the case where a
$\nu/\bar\nu$ separation of 90\% (70\%) is considered for $\mu$-like
($e$-like) events, as in Fig.~\ref{fig:disc}; the upper limit
corresponds to the case when the separation for $\mu$-like events is
reduced down to 70\%.

We would also like to point out that, once a 4\,MW 8\,GeV proton beam
becomes available from Project~X, muon cooling is added and the
detector mass is increased by a factor 1-3, the performance of this
facility reaches the $5^\circ$-level in CP accuracy, comparable to the
baseline NF. Therefore, neither the initial energy of 5\,GeV nor the
baseline do need to be changed in later stages.

In summary, using 10 times less useful muon decays and a 10 times
smaller detector mass with respect to the baseline NF still allows a
muon decay based neutrino beam to outperform realistic conventional
beams like LBNE phase I, while offering at the same time the path to a
full scale NF. Such low luminosity can be achieved using existing
proton drivers and without muon cooling. The reduced
muon energy makes it possible to use a shorter baseline around
1300\,km, and the use of a magnetized LAr detector allows to fully
exploit the physics potential of the platinum channel, which is
crucial for the overall performance of the facility. These choices for
baseline and detector technology ensure a good synergy with the first
stage of a superbeam program. The big open question is how well can a
LAr detector perform this task.

This work has been supported by the U.S. Department of Energy under
award number \protect{DE-SC0003915}. We would like to thank M.~Bishai
and M.~Bass for helping us to reproduce the LBNE sensitivities.


\begin{thebibliography}{27}
\expandafter\ifx\csname natexlab\endcsname\relax\def\natexlab#1{#1}\fi
\expandafter\ifx\csname bibnamefont\endcsname\relax
  \def\bibnamefont#1{#1}\fi
\expandafter\ifx\csname bibfnamefont\endcsname\relax
  \def\bibfnamefont#1{#1}\fi
\expandafter\ifx\csname citenamefont\endcsname\relax
  \def\citenamefont#1{#1}\fi
\expandafter\ifx\csname url\endcsname\relax
  \def\url#1{\texttt{#1}}\fi
\expandafter\ifx\csname urlprefix\endcsname\relax\def\urlprefix{URL }\fi
\providecommand{\bibinfo}[2]{#2}
\providecommand{\eprint}[2][]{\url{#2}}

\bibitem[{\citenamefont{An et~al.}(2013)}]{An:2012bu}
\bibinfo{author}{\bibfnamefont{F.}~\bibnamefont{An}} \bibnamefont{et~al.}
  (\bibinfo{collaboration}{Daya Bay Collaboration}), \bibinfo{journal}{Chin.
  Phys.} \textbf{\bibinfo{volume}{C37}}, \bibinfo{pages}{011001}
  (\bibinfo{year}{2013}), \eprint{1210.6327}.

\bibitem[{\citenamefont{Ahn et~al.}(2012)}]{Ahn:2012nd}
\bibinfo{author}{\bibfnamefont{J.}~\bibnamefont{Ahn}} \bibnamefont{et~al.}
  (\bibinfo{collaboration}{RENO collaboration}),
  \bibinfo{journal}{Phys.Rev.Lett.} \textbf{\bibinfo{volume}{108}},
  \bibinfo{pages}{191802} (\bibinfo{year}{2012}), \eprint{1204.0626}.

\bibitem[{\citenamefont{Abe et~al.}(2012)}]{Abe:2012tg}
\bibinfo{author}{\bibfnamefont{Y.}~\bibnamefont{Abe}} \bibnamefont{et~al.}
  (\bibinfo{collaboration}{Double Chooz Collaboration}),
  \bibinfo{journal}{Phys.Rev.} \textbf{\bibinfo{volume}{D86}},
  \bibinfo{pages}{052008} (\bibinfo{year}{2012}), \eprint{1207.6632}.

\bibitem[{\citenamefont{Huber et~al.}(2009)\citenamefont{Huber, Lindner,
  Schwetz, and Winter}}]{Huber:2009cw}
\bibinfo{author}{\bibfnamefont{P.}~\bibnamefont{Huber}},
  \bibinfo{author}{\bibfnamefont{M.}~\bibnamefont{Lindner}},
  \bibinfo{author}{\bibfnamefont{T.}~\bibnamefont{Schwetz}}, \bibnamefont{and}
  \bibinfo{author}{\bibfnamefont{W.}~\bibnamefont{Winter}},
  \bibinfo{journal}{JHEP} \textbf{\bibinfo{volume}{0911}}, \bibinfo{pages}{044}
  (\bibinfo{year}{2009}), \eprint{0907.1896}.

\bibitem[{\citenamefont{Agarwalla et~al.}(2012)\citenamefont{Agarwalla,
  Akhmedov, Blennow, Coloma, Donini et~al.}}]{Agarwalla:2012mz}
\bibinfo{author}{\bibfnamefont{S.}~\bibnamefont{Agarwalla}},
  \bibinfo{author}{\bibfnamefont{E.}~\bibnamefont{Akhmedov}},
  \bibinfo{author}{\bibfnamefont{M.}~\bibnamefont{Blennow}},
  \bibinfo{author}{\bibfnamefont{P.}~\bibnamefont{Coloma}},
  \bibinfo{author}{\bibfnamefont{A.}~\bibnamefont{Donini}},
  \bibnamefont{et~al.} (\bibinfo{year}{2012}), \eprint{1209.2825}.

\bibitem[{CDR()}]{CDR}
\bibinfo{note}{LBNE Conceptual Design Report from Oct 2012, volume 1},
  \urlprefix\url{https://sharepoint.fnal.gov/project/lbne/LBNE%20at%20Work/Sit%
ePages/Reports%20and%20Documents.aspx}.

\bibitem[{\citenamefont{Akiri et~al.}(2011)}]{Akiri:2011dv}
\bibinfo{author}{\bibfnamefont{T.}~\bibnamefont{Akiri}} \bibnamefont{et~al.}
  (\bibinfo{collaboration}{LBNE Collaboration}) (\bibinfo{year}{2011}),
  \eprint{1110.6249}.

\bibitem[{\citenamefont{Huber et~al.}(2008)\citenamefont{Huber, Mezzetto, and
  Schwetz}}]{Huber:2007em}
\bibinfo{author}{\bibfnamefont{P.}~\bibnamefont{Huber}},
  \bibinfo{author}{\bibfnamefont{M.}~\bibnamefont{Mezzetto}}, \bibnamefont{and}
  \bibinfo{author}{\bibfnamefont{T.}~\bibnamefont{Schwetz}},
  \bibinfo{journal}{JHEP} \textbf{\bibinfo{volume}{0803}}, \bibinfo{pages}{021}
  (\bibinfo{year}{2008}), \eprint{0711.2950}.

\bibitem[{\citenamefont{Coloma et~al.}(2012{\natexlab{a}})\citenamefont{Coloma,
  Huber, Kopp, and Winter}}]{Coloma:2012ji}
\bibinfo{author}{\bibfnamefont{P.}~\bibnamefont{Coloma}},
  \bibinfo{author}{\bibfnamefont{P.}~\bibnamefont{Huber}},
  \bibinfo{author}{\bibfnamefont{J.}~\bibnamefont{Kopp}}, \bibnamefont{and}
  \bibinfo{author}{\bibfnamefont{W.}~\bibnamefont{Winter}}
  (\bibinfo{year}{2012}{\natexlab{a}}), \eprint{1209.5973}.

\bibitem[{\citenamefont{Coloma et~al.}(2012{\natexlab{b}})\citenamefont{Coloma,
  Donini, Fernandez-Martinez, and Hernandez}}]{Coloma:2012wq}
\bibinfo{author}{\bibfnamefont{P.}~\bibnamefont{Coloma}},
  \bibinfo{author}{\bibfnamefont{A.}~\bibnamefont{Donini}},
  \bibinfo{author}{\bibfnamefont{E.}~\bibnamefont{Fernandez-Martinez}},
  \bibnamefont{and}
  \bibinfo{author}{\bibfnamefont{P.}~\bibnamefont{Hernandez}},
  \bibinfo{journal}{JHEP} \textbf{\bibinfo{volume}{1206}}, \bibinfo{pages}{073}
  (\bibinfo{year}{2012}{\natexlab{b}}), \eprint{1203.5651}.

\bibitem[{\citenamefont{Geer}(1998)}]{Geer:1997iz}
\bibinfo{author}{\bibfnamefont{S.}~\bibnamefont{Geer}},
  \bibinfo{journal}{Phys.Rev.} \textbf{\bibinfo{volume}{D57}},
  \bibinfo{pages}{6989} (\bibinfo{year}{1998}), \eprint{hep-ph/9712290}.

\bibitem[{\citenamefont{Cervera et~al.}(2000)\citenamefont{Cervera, Donini,
  Gavela, Gomez~Cadenas, Hernandez et~al.}}]{Cervera:2000kp}
\bibinfo{author}{\bibfnamefont{A.}~\bibnamefont{Cervera}},
  \bibinfo{author}{\bibfnamefont{A.}~\bibnamefont{Donini}},
  \bibinfo{author}{\bibfnamefont{M.}~\bibnamefont{Gavela}},
  \bibinfo{author}{\bibfnamefont{J.}~\bibnamefont{Gomez~Cadenas}},
  \bibinfo{author}{\bibfnamefont{P.}~\bibnamefont{Hernandez}},
  \bibnamefont{et~al.}, \bibinfo{journal}{Nucl.Phys.}
  \textbf{\bibinfo{volume}{B579}}, \bibinfo{pages}{17} (\bibinfo{year}{2000}),
  \eprint{hep-ph/0002108}.

\bibitem[{\citenamefont{Huber et~al.}(2006)\citenamefont{Huber, Lindner,
  Rolinec, and Winter}}]{Huber:2006wb}
\bibinfo{author}{\bibfnamefont{P.}~\bibnamefont{Huber}},
  \bibinfo{author}{\bibfnamefont{M.}~\bibnamefont{Lindner}},
  \bibinfo{author}{\bibfnamefont{M.}~\bibnamefont{Rolinec}}, \bibnamefont{and}
  \bibinfo{author}{\bibfnamefont{W.}~\bibnamefont{Winter}},
  \bibinfo{journal}{Phys.Rev.} \textbf{\bibinfo{volume}{D74}},
  \bibinfo{pages}{073003} (\bibinfo{year}{2006}), \eprint{hep-ph/0606119}.

\bibitem[{\citenamefont{Donini et~al.}(2002)\citenamefont{Donini, Meloni, and
  Migliozzi}}]{Donini:2002rm}
\bibinfo{author}{\bibfnamefont{A.}~\bibnamefont{Donini}},
  \bibinfo{author}{\bibfnamefont{D.}~\bibnamefont{Meloni}}, \bibnamefont{and}
  \bibinfo{author}{\bibfnamefont{P.}~\bibnamefont{Migliozzi}},
  \bibinfo{journal}{Nucl.Phys.} \textbf{\bibinfo{volume}{B646}},
  \bibinfo{pages}{321} (\bibinfo{year}{2002}), \eprint{hep-ph/0206034}.

\bibitem[{\citenamefont{Choubey et~al.}(2011)}]{NF:2011aa}
\bibinfo{author}{\bibfnamefont{S.}~\bibnamefont{Choubey}} \bibnamefont{et~al.}
  (\bibinfo{collaboration}{IDS-NF Collaboration}) (\bibinfo{year}{2011}),
  \eprint{1112.2853}.

\bibitem[{\citenamefont{Kyberd et~al.}(2012)}]{Kyberd:2012iz}
\bibinfo{author}{\bibfnamefont{P.}~\bibnamefont{Kyberd}} \bibnamefont{et~al.}
  (\bibinfo{collaboration}{nuSTORM Collaboration}) (\bibinfo{year}{2012}),
  \eprint{1206.0294}.

\bibitem[{\citenamefont{Ballett and Pascoli}(2012)}]{Ballett:2012rz}
\bibinfo{author}{\bibfnamefont{P.}~\bibnamefont{Ballett}} \bibnamefont{and}
  \bibinfo{author}{\bibfnamefont{S.}~\bibnamefont{Pascoli}},
  \bibinfo{journal}{Phys.Rev.} \textbf{\bibinfo{volume}{D86}},
  \bibinfo{pages}{053002} (\bibinfo{year}{2012}), \eprint{1201.6299}.

\bibitem[{\citenamefont{Tang and Winter}(2010)}]{Tang:2009wp}
\bibinfo{author}{\bibfnamefont{J.}~\bibnamefont{Tang}} \bibnamefont{and}
  \bibinfo{author}{\bibfnamefont{W.}~\bibnamefont{Winter}},
  \bibinfo{journal}{Phys.Rev.} \textbf{\bibinfo{volume}{D81}},
  \bibinfo{pages}{033005} (\bibinfo{year}{2010}), \eprint{0911.5052}.

\bibitem[{\citenamefont{\protect{Muon Accelerator Program}}(2013)}]{MAP}
\bibinfo{author}{\bibnamefont{\protect{Muon Accelerator Program}}},
  \emph{\bibinfo{title}{Muon accelerator program snowmass whitepaper}}
  (\bibinfo{year}{2013}), \bibinfo{note}{in preparation}.

\bibitem[{\citenamefont{Huber et~al.}(2002)\citenamefont{Huber, Lindner, and
  Winter}}]{Huber:2002mx}
\bibinfo{author}{\bibfnamefont{P.}~\bibnamefont{Huber}},
  \bibinfo{author}{\bibfnamefont{M.}~\bibnamefont{Lindner}}, \bibnamefont{and}
  \bibinfo{author}{\bibfnamefont{W.}~\bibnamefont{Winter}},
  \bibinfo{journal}{Nucl.Phys.} \textbf{\bibinfo{volume}{B645}},
  \bibinfo{pages}{3} (\bibinfo{year}{2002}), \eprint{hep-ph/0204352}.

\bibitem[{\citenamefont{Bross et~al.}(2008)\citenamefont{Bross, Ellis, Geer,
  Mena, and Pascoli}}]{Bross:2007ts}
\bibinfo{author}{\bibfnamefont{A.~D.} \bibnamefont{Bross}},
  \bibinfo{author}{\bibfnamefont{M.}~\bibnamefont{Ellis}},
  \bibinfo{author}{\bibfnamefont{S.}~\bibnamefont{Geer}},
  \bibinfo{author}{\bibfnamefont{O.}~\bibnamefont{Mena}}, \bibnamefont{and}
  \bibinfo{author}{\bibfnamefont{S.}~\bibnamefont{Pascoli}},
  \bibinfo{journal}{Phys.Rev.} \textbf{\bibinfo{volume}{D77}},
  \bibinfo{pages}{093012} (\bibinfo{year}{2008}), \eprint{0709.3889}.

\bibitem[{\citenamefont{Kyberd et~al.}(2009)\citenamefont{Kyberd, Ellis, Bross,
  Geer, Mena et~al.}}]{Bross:2009zzb}
\bibinfo{author}{\bibfnamefont{P.}~\bibnamefont{Kyberd}},
  \bibinfo{author}{\bibfnamefont{M.}~\bibnamefont{Ellis}},
  \bibinfo{author}{\bibfnamefont{A.}~\bibnamefont{Bross}},
  \bibinfo{author}{\bibfnamefont{S.}~\bibnamefont{Geer}},
  \bibinfo{author}{\bibfnamefont{O.}~\bibnamefont{Mena}}, \bibnamefont{et~al.}
  (\bibinfo{year}{2009}), \bibinfo{note}{{FERMILAB-FN-0836-APC}}.

\bibitem[{\citenamefont{Fernandez~Martinez
  et~al.}(2010)\citenamefont{Fernandez~Martinez, Li, Pascoli, and
  Mena}}]{FernandezMartinez:2010zza}
\bibinfo{author}{\bibfnamefont{E.}~\bibnamefont{Fernandez~Martinez}},
  \bibinfo{author}{\bibfnamefont{T.}~\bibnamefont{Li}},
  \bibinfo{author}{\bibfnamefont{S.}~\bibnamefont{Pascoli}}, \bibnamefont{and}
  \bibinfo{author}{\bibfnamefont{O.}~\bibnamefont{Mena}},
  \bibinfo{journal}{Phys.Rev.} \textbf{\bibinfo{volume}{D81}},
  \bibinfo{pages}{073010} (\bibinfo{year}{2010}), \eprint{0911.3776}.

\bibitem[{\citenamefont{Indumathi and Sinha}(2009)}]{Indumathi:2009hg}
\bibinfo{author}{\bibfnamefont{D.}~\bibnamefont{Indumathi}} \bibnamefont{and}
  \bibinfo{author}{\bibfnamefont{N.}~\bibnamefont{Sinha}},
  \bibinfo{journal}{Phys.Rev.} \textbf{\bibinfo{volume}{D80}},
  \bibinfo{pages}{113012} (\bibinfo{year}{2009}), \eprint{0910.2020}.

\bibitem[{\citenamefont{Donini et~al.}(2011)\citenamefont{Donini,
  Gomez~Cadenas, and Meloni}}]{Donini:2010xk}
\bibinfo{author}{\bibfnamefont{A.}~\bibnamefont{Donini}},
  \bibinfo{author}{\bibfnamefont{J.}~\bibnamefont{Gomez~Cadenas}},
  \bibnamefont{and} \bibinfo{author}{\bibfnamefont{D.}~\bibnamefont{Meloni}},
  \bibinfo{journal}{JHEP} \textbf{\bibinfo{volume}{1102}}, \bibinfo{pages}{095}
  (\bibinfo{year}{2011}), \eprint{1005.2275}.

\bibitem[{\citenamefont{Dutta et~al.}(2012)\citenamefont{Dutta, Indumathi, and
  Sinha}}]{Dutta:2011mc}
\bibinfo{author}{\bibfnamefont{R.}~\bibnamefont{Dutta}},
  \bibinfo{author}{\bibfnamefont{D.}~\bibnamefont{Indumathi}},
  \bibnamefont{and} \bibinfo{author}{\bibfnamefont{N.}~\bibnamefont{Sinha}},
  \bibinfo{journal}{Phys.Rev.} \textbf{\bibinfo{volume}{D85}},
  \bibinfo{pages}{013003} (\bibinfo{year}{2012}), \eprint{1103.5578}.

\bibitem[{\citenamefont{Huber and Schwetz}(2008)}]{Huber:2008yx}
\bibinfo{author}{\bibfnamefont{P.}~\bibnamefont{Huber}} \bibnamefont{and}
  \bibinfo{author}{\bibfnamefont{T.}~\bibnamefont{Schwetz}},
  \bibinfo{journal}{Phys.Lett.} \textbf{\bibinfo{volume}{B669}},
  \bibinfo{pages}{294} (\bibinfo{year}{2008}), \eprint{0805.2019}.

\end{thebibliography}

\end{document}